\documentstyle[amssymb,aps,multicol,epsf]{revtex}

\begin{document}
\draft
\title{Potts model on complex networks}
\author{S.N. Dorogovtsev$^{1,3,\ast}$, A.V. Goltsev$^{2,3,\dagger}$, and J.F.F.
Mendes$^{2,\ddagger}$}
\address{
$^1$ Departamento de F\'{\i }sica and Centro de F\'{\i }sica do Porto,
Faculdade de Ci\^{e}ncias, Universidade do Porto,
Rua do Campo Alegre 687, 4169-007 Porto, Portugal\\
$^{2}$ Departamento de F\'{\i}sica, Universidade,
de Aveiro, Campus Universit\'{a}rio de Santiago, 3810-193 Aveiro, Portugal\\
$^{3}$ A.F. Ioffe Physico-Technical Institute, 194021 St. Petersburg, Russia}
\maketitle

\begin{abstract}
We consider the general $p$-state Potts model on random networks with a
given degree distribution (random Bethe lattices). We find the effect of the
suppression of a first order phase transition in this model when the degree
distribution of the network is fat-tailed, that is, in more precise terms,
when the second moment of the distribution diverges. In this situation the
transition is continuous and of infinite order, and size effect is
anomalously strong. In particular, in the case of $p=1$, we arrive at the
exact solution, which coincides with the known solution of the percolation
problem on these networks.
\end{abstract}


\begin{multicols}{2}
\narrowtext

%%%%%%%%%%%%%%%%%%%%%%%%%%%%%%%%%%%%%%%%%%%%%%%%%%%%%%%%%%%%%%%%%%%%%%% 

\section{Introduction}

Complex networks display a spectrum of unique effects \cite
{ba99,s01,ab01a,dm01c,n03,w99,dm03,ajb00d,asbs00}. Cooperative phenomena in
complex networks are attracting much attention these days. The theoretical
study of various cooperative models on random networks \cite
{ahs01,dgm02,lvvz02,g00,bw00,khhjmc01,hkc02,b02,ceah00a,cnsw00,cbh02,pv01,gdm03,it02}
has demonstrated that their critical behaviour is extremely far from that on
regular lattices and on `planar graphs' \cite{k86}.

In this paper, we report our exact and asymptotically exact results for the
thermodynamic properties of the $p$-state Potts model on uncorrelated random
networks with a given degree distribution. These networks are the undirected
graphs, maximally random (i.e., with maximum entropy) under the constraint that
their degree distribution is a given one, $P(k)$. Here, {\em degree} is the
number of connections of a vertex. Correlations between degrees of vertices
in such graphs are absent, as well as clustering. In graph theory, these
networks are called `labeled random graphs with a given degree sequence' or
`the configuration model' \cite{bbk72}. One should stress that this is the
minimal model of complex networks. Most of results on cooperative models on
networks were obtained just for this basic construction \cite
{ceah00a,cnsw00,cbh02,pv01,it02,mr95,nsw00,dm01e,pv02}. See, however, Refs. 
\cite{chk01,dms01f,kkkr02a,bb03}, where the Berezinskii-Kosterlitz-Thouless
percolation phase transition was studied in growing networks. See also Ref.~ 
\cite{vbmpv03} and references therein for cooperative models on networks
with correlations between degrees of the nearest-neighbour vertices.

The percolation problem on uncorrelated networks with a given degree
sequence has been studied in Refs.~\cite{ceah00a,cnsw00,cbh02}. The Ising
model on these networks was studied by simulations in Ref.~\cite{ahs01} and
solved in Refs.~\cite{dgm02,lvvz02} (see also Ref.~\cite{b02}). It was shown
that the presence of fat tails in the degree distributions of networks
dramatically changes the critical behaviour of these models. But the site
percolation problem and the Ising model are only particular cases of the
general $p$-state Potts model \cite{wu82,bbook82}. The site percolation is
equivalent to the one-state Potts model \cite{KF}, and the Ising model is
exactly the two-state Potts model. At $p\geq 3$, the Potts model shows
features, very different from those at $p=1$ and $p=2$. In standard
mean-field theory, the $p\!\geq\! 3$\,-Potts model has a first-order phase
transition in contrast to percolation and the Ising model, where the phase
transitions are of second order.

Thus, the Potts model provides an essentially more wide range of behaviours
than the percolation problem and the Ising model. The Potts model is related
to a number of outstanding problems in statistical and mathematical physics,
and in graph theory, e.g., the colouring graph problem, etc. (for numerous
applications of the Potts model see Refs.~\cite{wu82,bbook82}). In short,
this a basic model of statistical mechanics, the direct generalization of
the percolation problem and the Ising model. Consequently, before more
complicated cooperative models on networks, one has to solve the Potts model
on the simplest complex networks.

In this paper we present in a detailed form our results which we have
announced in Ref.~\cite{gdm03}. We demonstrate a strong effect of a fat tail
in the degree distribution of a network on a first-order phase transition
which occurs in the Potts model with $p\geqslant 3$. 
%%on a random graph with a given degree distribution. 
We observe the suppression of the first-order
phase transition in networks with a fat-tailed degree distribution.

It is convenient to use the power-law degree distribution $P(k)\propto
k^{-\gamma }$ for parametrization. Then, $\langle k^4\rangle $ diverges for $%
\gamma \leq 5$, $\langle k^3\rangle $ diverges for $\gamma \leq 4$, and $%
\langle k^2\rangle $ diverges for $\gamma \leq 3$.

We find that if $p \geq 3$, the first order phase transition occurs only for $\gamma >3$,
while for $\gamma \leq 3$ the phase transition is continuous. At $\gamma =3$
, the phase transition is of infinite order and similar to the
transition in the Ising model on these networks \cite{dgm02}. We also obtain
the exact solution of the one-state Potts model on networks and show
that it agrees with the solution of the percolation problem on networks \cite
{cbh02}.

In this paper we consider the large networks that have a tree-like `local
structure'. In other words, any finite environment of a vertex in the
infinite network looks like a tree. Interactions are transmitted through
edges, from vertex to vertex. Thus, it is the distribution of the number of
connections of a nearest neighbour of a vertex that is crucial. In the
uncorrelated random networks, this distribution is $kP(k)/\langle k\rangle $%
. Then the nearest neighbours of a vertex have the average number of
connections (the average degree) $\langle k^{2}\rangle /\langle k\rangle $,
its second nearest neighbours have the same average degree, and so on. This
value is greater than the average number of connections for the entire
network $\langle k\rangle $, and it is much greater than $\langle k\rangle $
if $\langle k^{2}\rangle $ is large. It is not the mean degree of a network
that determines its cooperative behaviour but, rather, the average degree of
the nearest neighbour of a vertex \cite{dgm02}. This enables us to estimate a
certain characteristic temperature of the $p$-state Potts model on the
network using the formula $T_{c}/J=1/\ln [(q+p-2)/(q-2)]$ for the $p$%
-state Potts model on a regular Cayley tree with the coordination number 
$q=\langle k^{2}\rangle /\langle k\rangle $ \cite{bbook82}. The result is

\begin{equation}
\frac J{T_c}=\ln \frac{\langle k^2\rangle +(p-2)\left\langle k\right\rangle 
}{\langle k^2\rangle -2\langle k\rangle } \,.  \label{Tc}
\end{equation}

This naive estimate is exact for the Ising model \cite{dgm02}. It also gives
the exact percolation threshold as we will show below. Note, however, that
the meaning of the critical temperature $T_{c}$ is different for $p=1,$2 and 
$p\geqslant 3$. At $p=1$ the parameter $q_{c}=\exp (J/T_{c})-1$ determines
the percolation threshold on networks ($q$ is the probability that a vertex
is retained). At $p=2,$ $T_{c}$ is the critical temperature of the
continuous phase transitions in the Ising model on networks (with $J$
replaced by $2J$). We shall show that for $p\geqslant 3$ and $\gamma >3$, the 
$p$-state Potts model undergoes a first order phase transition, and $%
T_{c}$ given by Eq.~\ref{Tc} is the low-temperature boundary of the region
of hysteresis phenomena. We shall show, however, that if $\gamma \leqslant
3$ and $p\geqslant 1$, the transition in the Potts model becomes continuous, 
and $%
T_{c}$ is again the critical temperature of the transition (see discussion
below).

\section{General approach}

Consider the $p$-state Potts model with the Hamiltonian: 
\begin{equation}
{\cal H}=-J\sum_{\langle ij\rangle }\delta _{\alpha _{i},\alpha
_{j}}-H\sum_{i}\delta _{\alpha _{i},1}\,,  \label{Hamilt}
\end{equation}
where the first sum is over all edges of the graph, the second one is over
all vertices. $\delta _{\alpha ,\beta }=0,1$ if $\alpha \neq \beta $ and $%
\alpha =\beta $, respectively. Each vertex $i$ can be in any of $p$ states,
i.e., $\alpha _{i}=1,2,\ldots,p$. We assume a `ferromagnetic' interaction between
the nearest-neighbouring vertices, i.e., $J>0.$ The `magnetic field' $H>0$
distinguishes the state $\alpha =1$. Hereafter, we set $J=1$. The $p$%
-state Potts model on the regular Cayley tree is solved exactly by
using recurrence relations \cite{bbook82}. As networks under discussion have
a local tree-like structure, we apply the method of recurrence relations to
our problem. Actually, we use the same approach as in our solution of the
Ising model on networks \cite{dgm02}.

Consider a vertex $0$ with $k_{0}$ adjacent vertices with `spins' in states $\alpha
_{1,i} $, $i=1,2,\ldots,k_{0}$. Due to the local tree-like structure, this vertex
may be treated as a root of a tree. We introduce

\begin{eqnarray}
&&g_{1,i}(\alpha _{0})=  \nonumber \\[5pt]
&&\sum_{\{\alpha _{l}\}}\exp \text{ }\!\![(\sum_{\langle lm\rangle }\,\delta
_{\alpha _{l},\alpha _{m}}+\delta _{\alpha _{0},\,\alpha
_{1,i}}+H\sum_{l}\delta _{\alpha _{l},1})/T]\,.  \label{gg}
\end{eqnarray}
The labels $l$ and $m$ run only over vertices that belong to sub-trees with
the root vertex $\alpha_{1,i}$, including this vertex. Then the partition
function is 
\begin{equation}
Z=\sum_{\alpha _{0}}e^{H\delta _{\alpha
_{0},1}/T}\prod_{i=1}^{k_{0}}g_{1,i}(\alpha _{0})\,.  \label{Z}
\end{equation}

Let 
\begin{equation}
x_{1,i}(\alpha )\equiv g_{1,i}(\alpha )/g_{1,i}(1)\,,  \label{x}
\end{equation}
then the `magnetic moment' $M$ of the vertex $0$ is 
\begin{eqnarray}
M &=&\frac{p}{p-1}\left\langle \delta _{\alpha _{0},1}-\frac{1}{p}%
\right\rangle _{T}  \nonumber \\[5pt]
&=&\frac{1}{p-1}\frac{(p-1)e^{H/T}-\sum_{\alpha \neq
1}\prod_{i=1}^{k_{0}}x_{1,i}(\alpha )}{e^{H/T}+\sum_{\alpha \neq
1}\prod_{i=1}^{k_{0}}x_{1,i}(\alpha )}\,,  \label{M}
\end{eqnarray}
where $\left\langle ...\right\rangle _{T}$ is the thermodynamic average. The
parameters $x_{1,i}$ describe the effects of the nearest neighbours on the
vertex $0$. In turn, $x_{1,i}$ are expressed in terms of parameters $%
x_{2,l}(\alpha )=g_{2,l}(\alpha )/g_{2,l}(1)$, $l=1,2,\ldots,k_{1,i}$, which
describe effects of vertices in the second shell on vertices in the first shell,
and so on. The following recurrence relation between $x_{n,j}(\alpha )$ and $%
x_{n+1,l}(\alpha )$ holds at $n$ $\geqslant 1$ and $\alpha \geqslant 2$: 
\begin{eqnarray}
&&x_{n,j}(\alpha )=  
\nonumber 
\\[5pt]
&&[e^{H/T}+e^{1/T}\prod_{l=1}^{k_{n,j}-1}x_{n+1,l}(\alpha )+\sum_{\beta \neq
1,\alpha }\prod_{l=1}^{k_{n,j}-1}x_{n+1,l}(\beta )]/  \nonumber \\[5pt]
&&[e^{(1+H)/T}+\sum_{\beta \neq 1}\prod_{l=1}^{k_{n,j}-1}x_{n+1,l}(\beta
)]\,.  \label{recurr1}
\end{eqnarray}

If a vertex $n+1,l$ is a dead end, then $x_{n+1,l}(\alpha )=1$ at all $%
\alpha $. Deriving the recurrence relations, we started from some vertex $0$
and then made the recurrence steps along sub-trees. While solving the
recurrence relations (\ref{recurr1}), we start from distant vertices, i.e.,
from large $n$, and descend along sub-trees to the vertex. Note that all
states with the index $\alpha \geqslant 2$ are equivalent in respect of
statistics. Only the state $\alpha =1$ is distinguished by the applied field $H$. 

The recurrence steps Eq.~(\ref{recurr1}) converge exponentially quickly to
the fixed point which does not depend on $\alpha $. This enables us to put $%
x_{n,j}(\alpha )=x_{n,j}$ in Eq.~(\ref{recurr1}) from the very beginning for the
sake of simplicity. Then Eqs.~(\ref{M}) and (\ref{recurr1}) take a form:

\begin{eqnarray}
M &=&\frac{e^{H/T}-\prod_{i=1}^{k_{0}}x_{1,i}}{e^{H/T}+(p-1)%
\prod_{i=1}^{k_{0}}x_{1,i}}\,,  \label{M2} \\[5pt]
x_{n,j} &=&y\left( \prod_{l=1}^{k_{n,j}-1}x_{n+1,l})\right) \,,
\label{recurr2}
\end{eqnarray}
where we introduce 
\begin{equation}
y(x)=\frac{e^{H/T}+(e^{1/T}+p-2)x}{e^{(1+H)/T}+(p-1)x}\,.  \label{y(x)}
\end{equation}
Note that at $H=0$, in the paramagnetic phase $x_{n,l}=1$ while at $H>0$, $%
x_{n,l}\leqslant 1$. We stress that Eqs.~(\ref{M2}) and (\ref{recurr2}) are
exact 
for any tree-like graph.

\section{Exact solution at $p=1$: percolation on networks}

The one-state Potts model on networks at $H=0$ is of a special interest,
as it relates to percolation on networks. This limiting case
can be solved exactly for an uncorrelated random graph with an arbitrary
degree distribution. For this let us consider the recurrence relation
(\ref{recurr2}) at $p=1$ and $H=0$: 
\begin{equation}
x_{n,j}=e^{-1/T}+(1-e^{-1/T})\prod_{l=1}^{k_{n,j}-1}x_{n+1,l}\,.
\label{recurr-p1}
\end{equation}
Note that in an uncorrelated random graph, the degrees $k_{n,j}$ are
independent random variables. Since $x_{n+1,l}$ in this equation depends
only on $k_{m,j}$ with $m\geqslant n+1$, one can average the left and right
sides of the equation over the ensemble of random graphs with a degree
distribution $P(k)$ and find the self-consistent equation for the average $%
\left\langle x_{n,j}\right\rangle $. In the limit $n\rightarrow \infty $,
the fixed point $\left\langle x_{n,j}\right\rangle \rightarrow x$ of the
exact recurrence relation is given by 
\begin{equation}
x=1-q+\frac{q}{\left\langle k\right\rangle }\sum_{k}P(k)kx^{k-1}\,,
\label{percolation}
\end{equation}
where we introduce the parameter 
\begin{equation}
q=1-e^{-1/T}\,.  \label{q}
\end{equation}
It is important to note that Eq.~(\ref{q}) establishes the relation between
the one-state Potts model and site percolation on uncorrelated networks.
The latter is described by Eq.~(\ref{percolation}) \cite{cbh02}. 
%%In this respect t
The parameter $q$ has the meaning of the retained fraction of
vertices. The critical temperature $T_{c}$, Eq.~(\ref{Tc}), at $p=1$, 
determines the percolation threshold $q_{c}=\left\langle k\right\rangle
/(\left\langle k^{2}\right\rangle -\left\langle k\right\rangle )$. The
strong influence of the fat tail of the degree distribution 
%%$P(k)$ 
on
percolation critical exponents has been revealed and studied in detail in
Ref.~\cite{cbh02}.

\section{First order phase transition}

It is well known that the $p$-state Potts model, in the framework of the
standard mean-field theory, undergoes a first order phase transition for all $%
p\geqslant 3$ \cite{wu82,kms54}. The general approach derived above enables
us to consider, at $H=0$, the influence of the fat tail of the degree distribution 
%%at $H=0$ 
on the transition.

At $p\geqslant 3$, in order to average over the ensemble of random graphs
with a given degree distribution we use the effective medium approach
developed in Ref.~\cite{dgm02} for the Ising model on networks. First at
all, as at $H\geqslant 0$ we have $x_{n,l}\leqslant 0$, it is convenient to
introduce positive parameters $h_{n,l}$: $x_{n,l}=\exp (-h_{n,l})$. They are
independent random parameters and may be considered as random effective
fields acting on a vertex in the $n$-th shell from neighbouring `spins' in the $%
n+1$-th shell. Then, Eqs.~(\ref{M2}) and (\ref{recurr2}) take the form:

\begin{eqnarray}
M &=&\frac{e^{H/T}-\exp (-\sum_{l=1}^{k_{0}}h_{1,l})}{e^{H/T}+(p-1)\exp
(-\sum_{l=1}^{k_{0}}h_{1,l})}\,,  \label{M3} \\[5pt]
\text{ }h_{n,j} &=&-\ln \left\{ y\left[ \exp \left(
-\sum_{l=1}^{k_{n,j}-1}h_{n+1,l}\right) \right] \right\} \,.  \label{recurr3}
\end{eqnarray}
At dead ends we have $h_{n+1,l}=0.$ At $H=0$ in the paramagnetic phase $%
h_{n,l}=0$, while in the ordered phase $h_{n,l}\neq 0.$ Equations (\ref{M3})
and (\ref{recurr3}) determine the magnetization $M$ of a graph as a function
of $T$ and $H$.

While solving the recurrence relations (\ref{recurr3}), we start from
distant spins with $h\approx 0$ and descend along sub-trees to $\alpha $. In
the limit $n\rightarrow \infty$, the parameter $h_{1,i}$ is the fixed point
of the recurrence steps. The thermodynamic behaviour is determined by this
fixed point.

The right-hand sides of Eqs. (\ref{M3}) and (\ref{recurr3}) depend only on
the sum of the independent and equivalent random variables $h_{n,j}$. So,
let us use the following ansatz\cite{dgm02}:

\begin{equation}
\sum_{l=1}^kh_{n,l}\approx kh+{\cal O}(k^{1/2})\,,  \label{approx}
\end{equation}
where $h\equiv \langle h_{n,l}\rangle $ is the average value of the
`effective field' acting on a vertex. This approximation takes into account
the most `dangerous' highly connected vertices in the best way. With this
ansatz, averaging over the ensemble of random graphs and applying the ansatz
(\ref{approx}) to Eqs.~(\ref{M3}) and (\ref{recurr3}), we obtain %%
\begin{eqnarray}
\left\langle M\right\rangle &=&\sum_kP(k)\frac{e^{H/T}-e^{-kh}}{%
e^{H/T}+(p-1)e^{-kh}} \,,  \label{M-av} \\[5pt]
h &=&-\left\langle k\right\rangle ^{-1}\sum_kP(k)k\ln y[e^{-(k-1)h}]\equiv
G(h) \,.  \label{h-av}
\end{eqnarray}
The parameter $h$ plays the role of the order parameter. At $H=0$, $h=0$
above $T_c$ and $h>0$ below $T_c$.

Let us describe the thermodynamic properties of the Potts model with $%
p\geqslant 3$ on the infinite networks at $H=0$. For this, one must solve
the equation of state (\ref{h-av}).

At first, we consider the case $\langle k^{4}\rangle <\infty $. The
character of the transition and a characteristic temperature may be found
from the analysis of the expansion of $G(h)$ over small $h$. If $%
p\geqslant 3$ it is enough to take into account only first two terms: $%
G(h)=g_{1}h+g_{2}h^{2}+...$ where 
\begin{eqnarray}
g_{1} &=&\frac{\left\langle k(k-1)\right\rangle (e^{1/T}-1)}{\left\langle
k\right\rangle (e^{1/T}+p-1)},  \label{coeff1} \\
g_{2} &=&\frac{\left\langle k(k-1)^{2}\right\rangle (e^{1/T}-1)(p-2)}{%
2\left\langle k\right\rangle (e^{1/T}+p-1)^{2}}.  \label{coeff2}
\end{eqnarray}
One sees that at high temperatures, the coefficient $g_{1}\approx 0.$ When
temperature decreases, $g_{1}$ increases and becomes larger than 1. The
point where $g_{1}=1$ is the specific point. The corresponding temperature $%
T_{c}$ is given by Eq.~(\ref{Tc}). As $g_{2}>0$ at all temperatures, the
Potts model with $p\geqslant 3$ undergoes a first order phase transition
from the paramagnetic phase to the ordered one. Note that for the Ising
model ($p=2)$ the coefficients $g_{2}=0$ and $g_{3}<0$, and the phase
transition is continuous\cite{dgm02}. In the paramagnetic phase the
probabilities to find any vertex in the states $\alpha =1,2,\ldots,p$ are equal,
which corresponds to the solution $h=0$. In the ordered phase, one of the
states, $\alpha =1$ in our consideration, has a larger probability in
comparison to other states with $\alpha \geqslant 2$. This corresponds to
the solution with $h>0$.

\begin{figure}
\epsfxsize=74mm
\centerline{\epsffile{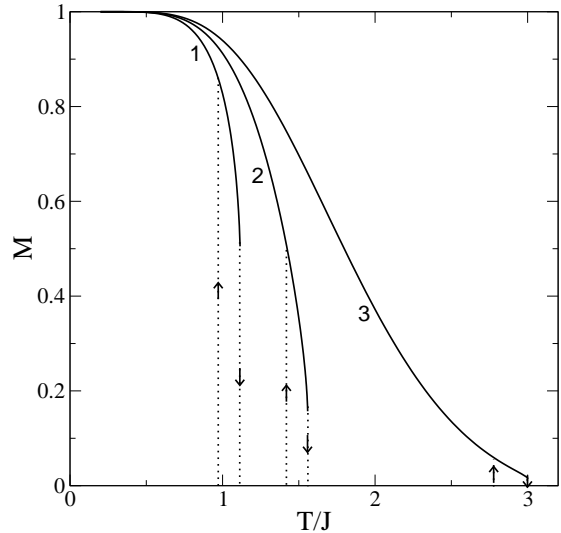}}
\caption{
The temperature dependence of the magnetization of the $p$-state Potts model with $p=5$ on uncorrelated random networks with the
degree distribution $P(k)=Ak^{-\gamma}$ for different exponents $\gamma$: 
1) $\gamma =6$; 2) $\gamma =4$; 
3) $\gamma =3.2$. The temperature region between dotted lines for each
curve is the hysteresis region. The up and down arrows correspond cooling
and heating, respectively.
} 
\label{f1}
\end{figure}

In order to find the temperature behaviour of the magnetization in the first
order phase transition, we have solved numerically the Eq.~(\ref{h-av}). At large $%
\gamma $, the system undergoes a standard first order phase transition (see
the temperature behaviour of $M$ in Fig.~\ref{f1} for $\gamma =6$). The
decrease of $\gamma $ modifies the second term in the expansion of $G(h)$:
at $\gamma =4$ we have $\langle k^{4}\rangle \rightarrow \infty $ and $%
G(h)=g_{1}h+g_{2}^{\prime }h^{2}\ln (1/h)+\ldots $, at $3<\gamma <4$ we find 
$G(h)=g_{1}h+g_{2}^{\prime }h^{\gamma -2}+\ldots $ In the range $3<\gamma
\leqslant 4$, the derivative $dG/dh$ at small $h$ remains to be positive and
the transition is still of the first order (see in Fig.~\ref{f1} our results
for $\gamma =4$ and 3.2). Results of the numerical solution of Eqs.~(\ref
{h-av}) and (\ref{M-av}) reveals that when $\gamma $ approaches 3 from above, 
the jump of the magnetic moment in the first order phase transition tends to
zero. Moreover, it is interesting to note that the temperature behaviour of $%
M $ in a wide temperature range just after the jump follows the exponential
law $M\sim \exp (-cT)$ with the constant $c$ which depends on the complete
degree distribution $P(k)$ (see the results in Fig.~\ref{f1} for $\gamma
=3.2)$. It is the behaviour that is expected for the infinite order phase
transition at $\gamma=3$ \cite{dgm02}.

%%%%%%%%%%%%%%%%%%%%%%%%%%%%%%%%%%%%%%%%%%%%%%%%%%%%%%%%%%%%%%%%%%%%%%%%%
%%%%%%%%%%%%%%%%%%%%%%%%%%%%%%%%%%%%%%%%%%%%%%%%%%%%%%%%%%%%%%%%%%%%%%%%%

%%%%%%%%%%%%%%%%%%%%%%%%%%%%%%%%%%%%%%%%%%%%%%%%%%%%%%%%%%%%%%%%%%%%%%%%%
%%%%%%%%%%%%%%%%%%%%%%%%%%%%%%%%%%%%%%%%%%%%%%%%%%%%%%%%%%%%%%%%%%%%%%%%%

In the case of the first order phase transition, there is a temperature
region, $T_{2}<T<T_{1}$, of hysteresis phenomena owing to the existence of
metastable states. For the Potts model, the low temperature boundary $T_{2}$
is the temperature below which the ordered state with $h\neq 0$ is the only
stable state. In other words, below $T_{2}$ the free energy as a
function of the order parameter has only one minimum corresponding to $%
h\neq 0$, while the solution $h=0$ corresponds to the maximum. One can show
that $T_{2}$ is determined by the equation $g_{1}=1$, i.e. $T_{2}=T_{c}$. In
the range $T_{2}<T<T_{1}$, the states with $h\neq 0$ and $h=0$ correspond to
the minimum of the free energy, i.e. one of these states, having larger
energy, is metastable. At temperatures $T>T_{1}$ the paramagnetic state, $%
h=0,$ is the only possible solution of the equation of state (\ref{h-av}). $%
T_{1}$ is determined by the set of Eqs.~(\ref{h-av}) and $dG(h)/dh=1.$ The
hysteresis region is shown in Fig.~\ref{f1} at different $\gamma $. One can
see that its width $\Delta T=T_{1}-T_{2}$ increases with increasing $\gamma $%
, however, $\Delta T/T_{2}\rightarrow 0.$

\section{Continuous transition for $2<\protect\gamma \leqslant 3$}

{\em The case} $\gamma=3$.---Here, the second moment $\left\langle
k^{2}\right\rangle $ diverges. Using, for brevity, the continuum
approximation for the degree distribution, we obtain $G(h)\approx
(\left\langle k\right\rangle h/(pT))\ln [p/(\langle k\rangle h)]$. One sees
that at small $h$, the second derivative $d^{2}G(h)/dh^{2}$ is negative in
contrast to the case $\gamma >3$ where $d^{2}G(h)/dh^{2}$ is positive. It
means the change of the order of the phase transition. Instead of the first
order phase transition discussed above, the $p$-state Potts model with $%
p\geqslant 3$ undergoes an infinite order phase transition at the critical
temperature $T_{c}$, Eq.~(\ref{Tc}), similarly to the Ising model \cite
{dgm02} and percolation \cite{cbh02}. This conclusion also agrees with the phenomenological theory of critical phenomena in complex networks \cite
{gdm03}.

One should emphasize that when $\left\langle k^{2}\right\rangle $ diverges,
the critical temperature $T_{c}$ is infinite for the infinite networks [see
Eq. (\ref{Tc})]. However, in any finite network, $\left\langle
k^{2}\right\rangle <\infty $, and $T_{c}$ is finite, although it may be very
high, $T_{c}\cong \langle k^{2}\rangle /(\langle k\rangle p)$ (see below).
At temperatures, which are much less than $T_{c}$, but where $h\ll 1$, so $%
T\gg 1$, we obtain

\begin{equation}
h\cong (p/\langle k\rangle )e^{-2pT/\left\langle k\right\rangle },\text{
\quad }M\cong e^{-2pT/\left\langle k\right\rangle },\text{ \quad }
\label{gamma3}
\end{equation}
Without the continuum approximation, we have, instead of $\langle k\rangle $
in the exponential, a constant which is determined by the complete form of $P(k)$.

{\it The case} $2<\gamma <3$.---Again $T_{c}$ for large networks is very high
and the phase transition is continuous. Using the expansion $G(h)\cong
g(\left\langle k\right\rangle /(pT))h^{\gamma -2}$, we find, in the range $%
1\ll T\ll T_{c}$, that the Potts model demonstrates the behaviour

\begin{equation}
h,\!M\sim T^{-1/(3-\gamma )}\!,  \label{gamma2-3}
\end{equation}
which is quite similar to the Ising model \cite{dgm02}.

In accordance with Eq. (\ref{Tc}), $T_{c}$ diverges when $\langle
k^{2}\rangle \rightarrow \infty $. However, in finite networks, $\langle
k^{2}\rangle $ is finite because of the finite-size cutoff of the degree
distribution. In scale-free networks, it is usually estimated as $%
k_{cut}\sim k_{0}N^{1/(\gamma -1)}$, where $N$ is the total number of
vertices in a network, $k_{0}$ is a `minimal degree' or the lower boundary
of the power-law dependence, and $\langle k\rangle \approx k_{0}(\gamma
-1)/(\gamma -2)$. Then, using estimates from Refs. \cite{dm01c,cbh02,pv02}
we obtain

\begin{eqnarray}
&&T_c\approx \frac{\langle k\rangle \ln N}p\ \ \ \ \ \ \ \ \ \ \ \ \ \ \ \ \
\ \ \ \ \ \ \ \ \ \ \ \ \ \ \mbox{at}\ \gamma =3 \,,  \nonumber \\[5pt]
&&T_c\approx \frac{4(\gamma -2)^2}{p(3-\gamma )(\gamma -1)}\langle k\rangle
N^{(3-\gamma )/(\gamma -1)}\ \ \mbox{for}\ 2<\gamma <3.  \label{2}
\end{eqnarray}
These expressions generalize the finite-size effect obtained for the Ising
model on networks\-\cite{dgm02}.

\section{Discussion}

Our investigations of the Potts and Ising models on uncorrelated random
networks demonstrate a general strong effect of fat tails in the degree
distribution on the order of the phase transition and its critical
temperature. In the Potts model with 
%%the number of 
$p\geqslant 3$ states,  
the phase transition of the first order occurs only when the second moment $%
\left\langle k^{2}\right\rangle $ is finite, i.e., at $\gamma >3$. When $%
\left\langle k^{2}\right\rangle $ diverges the behaviour of the Potts and
Ising models are similar. They undergoes the infinite order phase
transition. We suggest that this phenomenon 
(the suppression of a first order phase transition in favour of an infinite order one)  
is of a general nature and it
takes place in other cooperative models with short range interaction on
random networks. 

The phase diagram of the Potts model on uncorrelated networks was also
studied in Ref. \cite{it02} in the framework of a 
%%simplified 
simple mean field
approach. One should note that our results essentially differ from those in
Ref.~\cite{it02}. The reason for this difference is evidently the simplified
mean field theory of Ref. \cite{it02}.

S.N.D and J.F.F.M. were partially supported by the project
POCTI/MAT/46176/2002. A.G. acknowledges the support of the NATO program
OUTREACH. \newline

\noindent $^{\ast }$ E-mail address: sdorogov@fc.up.pt \newline
$^{\dagger }$ E-mail address: goltsev@mail.ioffe.ru \newline
$^{\ddagger }$ E-mail address: jfmendes@fis.ua.pt

\end{multicols}

\end{document}